\documentclass[aps,pre,reprint,twocolumn,superscriptaddress]{revtex4-1}
\usepackage{amsmath,amssymb,amsfonts,graphicx,bm}
\begin{document}

\title{Near-field thermodynamics and nanoscale energy harvesting}

\author{Ivan Latella}
\email{ilatella@ffn.ub.edu}
\affiliation{Departament de F\'{i}sica Fonamental, Facultat de F\'{i}sica, Universitat de Barcelona, Mart\'{i} i Franqu\`{e}s 1, 08028 Barcelona, Spain}

\author{Agust\'in P\'erez-Madrid}
\email{agustiperezmadrid@ub.edu}
\affiliation{Departament de F\'{i}sica Fonamental, Facultat de F\'{i}sica, Universitat de Barcelona, Mart\'{i} i Franqu\`{e}s 1, 08028 Barcelona, Spain}

\author{Luciano C. Lapas}
\email{luciano.lapas@unila.edu.br}
\affiliation{Universidade Federal da Integra\c{c}\~{a}o Latino-Americana, Caixa Postal 2067, 85867-970 Foz do Igua\c{c}u, Brazil}

\author{J. Miguel Rubi}
\email{mrubi@ub.edu}
\affiliation{Departament de F\'{i}sica Fonamental, Facultat de F\'{i}sica, Universitat de Barcelona, Mart\'{i} i Franqu\`{e}s 1, 08028 Barcelona, Spain}
\affiliation{Department of Chemistry, Imperial College London, SW7 2AZ London, United Kingdom
}


\begin{abstract}
We study the thermodynamics of near-field thermal radiation between two identical polar media at different temperatures. As an application, we consider an idealized energy harvesting process from sources at near room temperature at the nanoscale. We compute the maximum work flux that can be extracted from the radiation in the near-field regime and compare it with the corresponding maximum work flux in the blackbody regime. This work flux is considerably higher in the near-field regime. For materials that support surface phonon polaritons, explicit expressions for the work flux and an upper bound for the efficiency as functions of the surface wave frequency are obtained.
\end{abstract}


\newcommand{\dif}{\text{d}}
\newcommand{\ee}{\text{e}}
\newcommand{\ii}{\text{i}}
\newcommand{\kB}{k_\text{B}}
\newcommand{\sub}[1]{\text{#1}}
\newcommand{\vect}[1]{\bm{#1}}

\maketitle

\section{Introduction}

Near-field radiative heat transfer has recently attracted much attention due to the important role it plays in nanoscale physics and technology.
Thermal radiation heat transfer in this regime is considerably enhanced as compared to the blackbody limit~\cite{Rousseau:2009,Shen:2009,Shen:2012}.
That is, the amount of energy exchanged between bodies separated by submicron distances is notably higher than that for bodies separated by macroscopic distances~\cite{Perez-Madrid:2008,Perez-Madrid:2009,Perez-Madrid:2013}. The tunneling of evanescent electromagnetic waves is responsible for this enhancement, an effect that can rise only when the bodies are close to each other~\cite{Joulain:2005,Volokitin:2007}.
This phenomenon is exploited, for instance, in scanning thermal microscopy \cite{Majumdar:1999,Kittel:2005}, the recently developed near-field thermal transistor~\cite{Ben-Abdallah:2014}, and the generation of usable energy from thermal sources via thermophotovoltaic devices~\cite{Narayanaswamy:2003,Laroche:2006,Park:2008,Basu:2009,Messina:2013}.
Due to the contribution of evanescent modes, the local density of states is modified close to an interface separating two media~\cite{Joulain:2003}. This implies that the thermodynamic functions will also depend on this contribution~\cite{Dorofeyev:2011,Narayanaswamy:2013} and will show a very different behavior from the one shown in the far-field case.

Here we consider the thermodynamics of near-field thermal radiation and its application to energy harvesting. We focus on the radiation emitted by two identical polar media at different temperatures and compute the maximum work that can be extracted from this system. An upper bound for the efficiency is also discussed. Here we concentrate on the case where the temperature difference between the hot source and the receiver is small. In these conditions, on the one hand, converters working in the far field not only have low efficiencies, but also the power they supply is poor. This is due to the fact that converters in the far field require high temperature sources to operate in optimal conditions~\cite{Ilic:2012}. On the other hand, although the efficiency remains low if the temperature difference between the source and the receiver is small, the delivered power is notably higher for converters working in the near field. Thus, near-field radiation brings out the possibility for energy harvesting from 
sources of moderate temperature at the nanoscale.

\section{Fluctuating fields and energy flux}

Near-field radiative heat transfer is described using a semi-classical approach. That is, the classical Maxwell equations are utilized to describe the electromagnetic fields produced by currents in the material, while quantum statistics is introduced to account for the characteristic occupation numbers of photons in energy levels.
Such an approach is known as fluctuating electrodynamics~\cite{Rytov:1989}. 
The currents in the material are due to the random movement of charges associated to thermal excitations. That is, thermal radiation is produced by stochastic fluctuations of the currents.
Moreover, the Fourier components of the  electric and magnetic fields, $\vect{E}(\vect{r},\omega)$ and $\vect{H}(\vect{r},\omega)$, respectively, can be written in terms of the Fourier components of the current $\vect{j}(\vect{r},\omega)$ according to~\cite{Mulet:2002,Biehs:2012}
\begin{align}
&\vect{E}(\vect{r},\omega)=\ii\mu_0\omega\int_V\dif^3\vect{r}'\ \mathbb{G}^E(\vect{r},\vect{r}',\omega)\cdot \vect{j}(\vect{r}',\omega),\label{E}\\
&\vect{H}(\vect{r},\omega)=\int_V\dif^3\vect{r}'\ \mathbb{G}^H(\vect{r},\vect{r}',\omega)\cdot \vect{j}(\vect{r}',\omega),\label{H}
\end{align}
where $\mathbb{G}^E$ and $\mathbb{G}^H$ are the classical electric and magnetic Green tensors, respectively, and $\mu_0$ is the vacuum permeability.
The integrations extend over the volume $V$ in the material where the currents are present and these Green tensors relate the source current at point $\vect{r}'$ to the electric and magnetic fields at point $\vect{r}$ outside the volume.
Here we shall restrict ourselves to isotropic nonmagnetic materials.

In neutral materials, the statistical average of the fluctuating charge density and of the electric current density vanish.
Therefore, also the average value of the fluctuating electromagnetic fields is null.
However, although the average values of the fields vanish, their correlations are different from zero.
The associated correlation function for the Cartesian components $j_k$ ($k=1$, $2$, 3) of the current is given by the fluctuation-dissipation theorem~\cite{Polder:1971,Joulain:2005}
\begin{equation}
\begin{split}
\left\langle j_k(\vect{r},\omega)j_l^*(\vect{r}',\omega')\right\rangle&=4\pi\varepsilon_0\hbar\omega^2 n(\omega,T)\text{Im}\left[\varepsilon(\omega)\right]\\
&\qquad\times\delta_{kl}\delta(\omega-\omega')\delta(\vect{r}-\vect{r}'),
\end{split}
\label{FDT}
\end{equation}
where $\langle\,\cdots\rangle$ denotes statistical average, $\varepsilon_0$ is the permittivity of vacuum, $\hbar$ is the reduced Planck constant, the asterisk denotes the complex conjugate, and $\text{Im}\left[\varepsilon(\omega)\right]$ denotes the imaginary part of the complex dielectric constant of the medium $\varepsilon(\omega)$. Here
\begin{equation}
n(\omega,T)=\left[\ee^{\hbar\omega/(\kB T)}-1\right]^{-1}
\end{equation}
is the average number of photons in a single mode of frequency $\omega$ at equilibrium temperature $T$, and $\kB$ is the Boltzmann constant.
Therefore, given the Green tensors, the expressions for the fields (\ref{E}) and (\ref{H}), and the fluctuation-dissipation theorem (\ref{FDT}), the average of quantities like the Poynting vector $\vect{\Sigma}=\vect{E}\times\vect{H}$ can be computed (see Ref.~\cite{Joulain:2005}). The Poynting vector is quadratic in the fields and, hence, its average value does not vanish due to correlations.

Consider now two semi-infinite media with planar surfaces separated by a vacuum gap and with different temperatures; $T_1$ for medium 1 and $T_2$ for medium~2. The average of the component of the Poynting vector perpendicular to the surface of the material gives the energy flux (energy per unit time and surface) radiated by the fluctuating currents inside the material. Hence, the energy flux emitted by medium~1 and absorbed by the second medium can be written as 
\begin{equation}
\dot{U}(T_1)\equiv\left\langle \Sigma_z^{1\to2}\right\rangle=\left\langle \vect{E}_1\times \vect{H}_1\right\rangle\cdot \vect{e}_z,
\end{equation}
where $z$ denotes the direction perpendicular to the surface, and $\vect{e}_z$ is the unit vector in this direction.
On the other hand, the energy emitted by medium~2 and absorbed by medium~1 is given by $\dot{U}(T_2)=\left\langle \Sigma_z^{2\to1}\right\rangle$. Furthermore, for a given temperature $T$, the energy flux can be written as
\begin{equation}
\dot{U}(T)=\int_0^\infty\dif \omega\; \hbar\omega n(\omega,T)\varphi(\omega),
\label{energy_flux}
\end{equation}
where $\varphi(\omega)$ is the spectral flux of modes~\cite{Latella:2014}.
The net energy transfer $\Delta\dot{U}=\dot{U}(T_2)-\dot{U}(T_1)$ is therefore given by
\begin{equation}
\Delta\dot{U}=\int_0^\infty\dif \omega\; \hbar\omega \left[n(\omega,T_2)-n(\omega,T_1)\right]\varphi(\omega).
\end{equation}

The function $\varphi(\omega)$ is well-known for planar surfaces~\cite{Volokitin:2001,Mulet:2002,Joulain:2005,Volokitin:2007}.
Considering that the two materials are identical and introducing the reflection coefficients of the vacuum-material interface $R_\alpha(\kappa,\omega)$ for the polarizations $\alpha=\text{p},\text{s}$, the spectral flux of modes is given by
\begin{equation}
\begin{split}
\varphi(\omega)&=\sum_{\alpha=\text{p},\text{s}}\left\{\int_0^{\omega/c}\frac{\dif \kappa\, \kappa}{4\pi^2}\frac{\left[1-|R_\alpha(\kappa,\omega)|^2\right]^2}{\left|1-\ee^{2\ii\gamma d}R_\alpha^2(\kappa,\omega)\right|^2}\right.\\
&\left.\qquad\ +\int_{\omega/c}^\infty\frac{\dif \kappa\, \kappa}{\pi^2}\frac{\ee^{-2|\gamma|d}\text{Im}^2\left[ R_\alpha(\kappa,\omega)\right]}{\left|1-\ee^{-2|\gamma|d}R_\alpha^2(\kappa,\omega)\right|^2}\right\}.
\label{SFM}
\end{split}
\end{equation}
Here, $d$ is the separation between the surfaces, $c$ is the speed of light in vacuum, and $\kappa$ is the component of the wave vector parallel to the surfaces satisfying $\gamma=\left[(\omega/c)^2-\kappa^2\right]^{1/2}$. The quantity $\gamma$ is the component of the wave vector perpendicular to the surfaces in vacuum. This component of the wave vector in the media depends on the dielectric constant and takes the form $\gamma_\text{m}=\left[(\omega/c)^2\varepsilon-\kappa^2\right]^{1/2}$.
In addition, the reflection coefficients depend on the optical properties of the material under consideration and are given by~\cite{Joulain:2005}
\begin{equation}
R_\text{p}(\kappa,\omega)= \frac{\varepsilon\gamma-\gamma_\text{m}}{\varepsilon\gamma+\gamma_\text{m}},\qquad 
R_\text{s}(\kappa,\omega)= \frac{\gamma-\gamma_\text{m}}{\gamma+\gamma_\text{m}}.
\end{equation}

Two different contributions to the flux of modes can be identified in (\ref{SFM}): The first term in curly brackets accounts for propagative modes such that $\kappa<\omega/c$, while the second term corresponds to evanescent electromagnetic waves, where $\kappa>\omega/c$. Evanescent modes decay exponentially from the surface of the material, so that their contribution is negligible at relatively large distances. In addition, in order to contribute significantly, these modes have to lie in the range of excited thermal modes given by $n(\omega,T)$. Thermal radiation is thus enhanced if it is measured at distances from the surface much smaller than the thermal wavelength $\lambda_T=c\hbar/\kB T$, which for $T=300\,$K takes the value $\lambda_T=7.6\,\mu$m. Thus, at room temperature, these effects are appreciable at the nanoscale.  

\section{Entropy flux}

The entropy flux $\dot{S}$ (entropy per unit time and surface) associated to the radiation emitted by the surface of the material can be obtained from the energy flux by taking into account the usual thermodynamic relations. According to this, $\dot{S}$ must satisfy
\begin{equation}
\frac{1}{T}=\frac{\dif \dot{S}}{\dif \dot{U}}.
\end{equation}
This differential equation can be integrated to give
\begin{equation}
\dot{S}(T)=\int_0^T \dif T'\frac{1}{T'}\frac{\dif \dot{U}(T')}{\dif T'}. 
\label{entropy_flux}
\end{equation}
Assuming that the spectral flux of modes does not depend on $T$, the entropy flux (\ref{entropy_flux}) becomes~\cite{Latella:2014}
\begin{equation}
\dot{S}(T)=\int_0^\infty\dif \omega\; \kB  m(\omega,T) \varphi(\omega),
\label{Entropy}
\end{equation}
where we have introduced
\begin{equation}
\begin{split}
m(\omega,T)&= \left[1+n(\omega,T)\right]\ln\left[1+n(\omega,T)\right] \\
&\qquad-n(\omega,T)\ln n(\omega,T). 
\label{m}
\end{split}
\end{equation}

We want to stress that since, in general, the optical properties of the material can depend on temperature, $\varphi$ can also depend on temperature~\cite{Basu:2009}.
Nevertheless, if only variations of entropy flux are considered, (\ref{Entropy}) can still be used if $\varphi$ is constant or approximately constant in the considered working temperature range, and arbitrary otherwise (see Appendix).
We also emphasize that the expressions (\ref{energy_flux}) and (\ref{Entropy}) for the energy and entropy fluxes, respectively, rest on the assumption that the emitted radiation is in thermal equilibrium with the radiating body at temperature $T$.

\section{Thermodynamics of an ideal energy conversion process}
\label{ideal:conversion}

The conversion of thermal radiation can be seen as the evolution of the photon gas between two different states, in such a way that during its evolution the system delivers some useful work. Instead of having a cyclic machine, in this case the stationary energy and entropy fluxes lead to an stationary usable work flux. The first law of thermodynamics states that
\begin{equation}
\Delta \dot{U}+ \dot{Q}_\sub{e}+\dot{W}=0,
\label{first_law}
\end{equation}
where $\Delta \dot{U}$ is the variation of the internal energy of the photon gas, $\dot{Q}_\sub{e}$ is the heat flux delivered to the environment during the transformation and $\dot{W}$ is the flux of usable work that can be obtained during the process. Assuming that $\dot{Q}_\sub{e}$ is transferred isothermically to the environment at temperature $T_\sub{e}$, the variation of the entropy flux in the environment is given by $\Delta\dot{S}_\sub{e}=\dot{Q}_\sub{e}/T_\sub{e}$. Thus, $T_\sub{e}$ is also the temperature of the photon gas in its final state if we assume that it ends in thermal equilibrium with the environment. In addition, the second law of thermodynamics establishes that the variation of the entropy flux in the system $\Delta \dot{S}$ satisfies~\cite{deGroot:1984}
\begin{equation}
\Delta \dot{S}+\Delta \dot{S}_\sub{e}=\Delta \dot{S}_\sub{irr}\geq0, 
\label{second_law}
\end{equation}
where $\Delta \dot{S}_\sub{irr}$ is the entropy production that accounts for irreversibilities in the process of conversion. Combining (\ref{first_law}) and (\ref{second_law}), the work flux in the stationary regime can be written as
\begin{equation}
\dot{W}=T_\sub{e}\Delta \dot{S}-\Delta \dot{U}-T_\sub{e}\Delta \dot{S}_\sub{irr}.
\end{equation}

An ideal process is that for which no entropy is produced due to irreversibilities, $\Delta \dot{S}_\sub{irr}=0$. Thus, the ideal work flux $\dot{\mathcal{W}}$ delivered by the system during an ideal process is given by~\cite{Kjelstrup:2010}
\begin{equation}
\dot{\mathcal{W}}\equiv T_\sub{e}\Delta \dot{S}-\Delta \dot{U}. 
\label{ideal_work_flux}
\end{equation}
Of course, real energy conversion processes are not ideal.
However, even for a non-ideal process, $\dot{\mathcal{W}}$ gives relevant information about the energy conversion since it is the maximum work flux that can be obtained from the system for a given initial state. In our case, the initial state is defined by the temperature of the hotter radiating medium, which we denote by $T_\text{h}$, $T_\text{h}>T_\text{e}$. According to the previous discussion, the temperature of the second medium is $T_\text{e}$, so that $\Delta \dot{U}=\dot{U}(T_\sub{e})-\dot{U}(T_\sub{h})$, with the energy flux given by (\ref{energy_flux}). Furthermore, the variation of entropy flux is $\Delta \dot{S}=\dot{S}(T_\sub{e})-\dot{S}(T_\sub{h})$, and hence
\begin{equation}
\Delta \dot{S}=\int_0^\infty\dif \omega\; \kB  \left[m(\omega,T_\sub{e})-m(\omega,T_\sub{h}) \right]\varphi(\omega).
\label{delta_S}
\end{equation}

An important parameter that describes the mechanism of energy conversion is the efficiency of the process. For an ideal process, an adequate criterion that quantifies the performance of the conversion is given by the first law efficiency $\eta$. Hereafter we denote by $\bar{\eta}$ the efficiency of an ideal process. This can be written as~\cite{Kjelstrup:2010}
\begin{equation}
\bar{\eta}=\frac{\dot{\mathcal{W}}}{\dot{U}(T_\sub{h})},
\label{upper_bound}
\end{equation}
which is the ratio of the useful work flux to the input energy flux.
If the energy conversion process is non-ideal, that is $\Delta \dot{S}_\sub{irr}\neq0$, equation~(\ref{upper_bound}) still gives information about the process because $\bar{\eta}$ is an upper bound for the efficiency. Below we will consider $\bar{\eta}$ for the case of near-field thermal radiation energy conversion and compare it to the case of blackbody radiation.

\section{Near-field thermal energy harvesting}

Energy converters can be used with the purpose of capturing thermal energy from their surroundings and transform it into usable work. Here we consider this energy harvesting process in an ideal situation, which is a first approximation to the real case. We consider a semi-infinite medium acting as a thermal energy source at a temperature higher than the environment temperature. This medium could be a certain component of a device which, as a consequence of an independent task, is kept at a working temperature $T_\sub{h}$. A second semi-infinite medium is placed near the first one, with a vacuum gap separating the (planar) surfaces of the two media. The second medium is assumed to be in thermal equilibrium with the environment at temperature $T_\sub{e}$.
Due to the difference of temperatures, it is clear that a certain amount of work can be extracted from the thermal radiation. This function is assigned to the converter, which can be assumed to be coupled to the medium at temperature $T_\sub{e}$. The specific mechanism utilized by the converter to transform the radiation will determine the entropy production and, therefore, the efficiency of the process. If this mechanism is not particularized, bounds for the efficiency and work flux can be obtained by considering an ideal process, as discussed in Sec.~\ref{ideal:conversion}. These bounds will be obtained below. In particular, we will focus on the case where the difference of temperature between the hot medium and the environment is relatively small. Small temperature differences are the physically relevant situation for energy harvesters at the nanoscale, where near-field thermal radiation is the dominant contribution. In order to quantify the performance of the process in these conditions, we need suitable 
expressions for the maximum work flux and the upper bound for the efficiency $\bar{\eta}$.
Explicitly, a relatively small temperature difference $\Delta T=T_\sub{e}-T_\sub{h}$ is achieved if $|\Delta T|/T_0\ll1$, with $T_0=(T_\sub{e}+T_\sub{h})/2$.
The maximum usable work flux that an energy harvester can produce is given by (\ref{ideal_work_flux}) and, thus, using (\ref{energy_flux}) and (\ref{Entropy}), we expand (\ref{ideal_work_flux}) to leading order in $\Delta T$ and obtain
\begin{equation}
\dot{\mathcal{W}}=\frac{\kB (\Delta T)^2}{2T_0}\int_0^\infty\dif \omega\ \left(\frac{\hbar\omega}{\kB T_0}\right)^2\frac{\ee^{\hbar\omega/(\kB T_0)}\varphi(\omega)}{\left[\ee^{\hbar\omega/(\kB T_0)}-1\right]^2}.
\label{ideal_work_expanded}
\end{equation}
Notice that energetic and entropic contributions are equal to first order in $\Delta T$, therefore the leading contribution is of order $(\Delta T)^2$. Likewise, a suitable expression for $\bar{\eta}$ will be given below for small gap separations.

As we have seen, the energy transfer strongly depends on the optical properties of the emitters, which are introduced through $\varphi(\omega)$. 
In  what follows we will concentrate on the case of two identical polar media that support surface phonon polaritons, which are surface waves due to the coupling of phononic excitations with the electromagnetic fields~\cite{Mulet:2002,Joulain:2005}. If two planar sources supporting surface phonon polaritons are closely placed, these modes can be resonantly excited and thus produce a considerably increase of the emitted radiation~\cite{Joulain:2005,Volokitin:2007}.
Examples of materials that support these surface waves are, for instance, silicon carbide (SiC) and hexagonal boron nitride (hBN). The optical properties of these materials can be described by the Lorentz model~\cite{Palik:1998,Messina:2013}
\begin{equation}
\varepsilon(\omega)=\varepsilon_\infty\left(\frac{\omega^2_\sub{L}-\omega^2-\ii\Gamma\omega}{\omega^2_\sub{T}-\omega^2-\ii\Gamma\omega}\right), 
\label{Lorentz_model}
\end{equation}
where $\omega_\sub{L}$, $\omega_\sub{T}$, $\varepsilon_\infty$, and $\Gamma$ are characteristic parameters of the material. In addition, for these materials, the frequency of the surface phonon polariton of the single interface can be written as~\cite{Rousseau:2012}
\begin{equation}
\omega_0=\left(\frac{\varepsilon_\infty\omega_\sub{L}^2+\omega_\sub{T}^2}{\varepsilon_\infty+1}\right)^{1/2}.
\end{equation}
Furthermore, in the case of polar materials and when the gap width is small ($d\ll\lambda_T$), the dominant contribution in (\ref{SFM}) comes from p-polarized evanescent modes.
Retaining only this contribution, the spectral flux of modes in the near-field regime can be obtained using the near-monochromatic approximation and, hence, be written in terms of a Dirac $\delta$ distribution as~\cite{Latella:2014,Rousseau:2012}
\begin{equation}
\varphi_\sub{nf}(\omega)= g_d(\omega)\delta(\omega-\omega_0),
\label{SFM_nf}
\end{equation}
where
\begin{equation}
g_d(\omega)=\frac{\mathrm{Re}\left[\mathrm{Li}_2\left(R^2_\mathrm{p}(\omega)\right)\right]}{4\pi d^2f'(\omega)}.
\label{g_d}
\end{equation}
In Eq.~(\ref{g_d}), we have introduced
\begin{equation}
f(\omega)= \frac{\text{Im}\left[R_\text{p}^2(\omega)\right]}{\text{Im}^2\left[ R_\text{p}(\omega)\right]},
\end{equation}
with $f'(\omega)=\dif f(\omega)/\dif \omega$, and $\text{Li}_n(z)=\sum_{k=1}^\infty z^k/k^n$ is the polylogarithm function.
The electrostatic limit is also assumed in~(\ref{SFM_nf}), where the reflection coefficient does not depend on $\kappa$ and reads
\begin{equation}
R_\text{p}(\omega)=\frac{\varepsilon(\omega)-1}{\varepsilon(\omega)+1}.
\end{equation}
It is important to note that the function $g_d(\omega)$ is proportional to $1/d^2$ and, therefore, the thermodynamic quantities in the near-field are functions of the width of the vacuum gap separating the surfaces. We also emphasize that these arguments are valid for polar materials; for metallic surfaces, s-polarized fields dominate the heat transfer~\cite{Chapuis:2008}.

The expression (\ref{SFM_nf}) for the spectral flux of modes allows us to compute thermodynamic functions like~(\ref{ideal_work_expanded}) in the near-field regime. Indeed, to leading order in $\Delta T$ we have
\begin{equation}
\dot{\mathcal{W}}_\sub{nf}= \frac{\kB(\Delta T)^2}{2T_0} \left(\frac{\hbar\omega_0}{\kB T_0}\right)^2\frac{\ee^{\hbar\omega_0/(\kB T_0)}g_d(\omega_0)}{\left[\ee^{\hbar\omega_0/(\kB T_0)}-1\right]^2}.
\label{ideal_work_nf}
\end{equation}
In addition, expanding $\dot{U}(T_\sub{e})=\dot{U}(T_0-\Delta T/2)$ about $T_0$ from (\ref{energy_flux}) with (\ref{SFM_nf}), and taking (\ref{ideal_work_nf}) into account, the upper bound for the efficiency (\ref{upper_bound}) in the near-field becomes
\begin{equation}
\bar{\eta}_\text{nf}=\frac{\hbar\omega_0(\Delta T)^2}{2\kB T_0^3}\left[1-\ee^{-\hbar\omega_0/(\kB T_0)}\right]^{-1},
\label{eta_nf}
\end{equation}
where, again, we consider only the contribution to leading order in $\Delta T$. For arbitrary temperatures, $\bar{\eta}_\text{nf}$ correspond to the efficiency of near-monochromatic radiation~\cite{Latella:2014}.

To appreciate the performance of the energy harvesting process in the near-field, it is illuminating to compare (\ref{ideal_work_nf}) and (\ref{eta_nf}) with their analogues in the blackbody regime. In the blackbody regime, the corresponding spectral flux of modes $\varphi_\sub{bb}(\omega)=\omega^2/(2\pi c)^2$ is obtained by setting $R_\alpha=0$ in (\ref{SFM}).
Thus, the ideal work flux in this regime, $\dot{\mathcal{W}}_\sub{bb}$, is obtained from (\ref{ideal_work_flux}) using (\ref{energy_flux}) and (\ref{Entropy}) with $\varphi(\omega)=\varphi_\sub{bb}(\omega)$. Notice that using $\varphi(\omega)=\varphi_\sub{bb}(\omega)$ in (\ref{energy_flux}) leads to the Stefan-Boltzmann law. 
For small $\Delta T$ one obtains $\dot{\mathcal{W}}_\sub{bb}=2\sigma T_0^2(\Delta T)^2$, where $\sigma$ is the Stefan constant.
In addition, the bound for the efficiency in this case reads~\cite{Landsberg:1980}
\begin{equation}
\bar{\eta}_\sub{bb}= 1-\frac{4}{3}\frac{T_\sub{e}}{T_\sub{h}}+\frac{1}{3}\left(\frac{T_\sub{e}}{T_\sub{h}}\right)^4. 
\end{equation}

We now define
\begin{equation}
\Phi_0(d,T_\sub{e})\equiv\lim_{T_\sub{h}\to T_\sub{e}}\frac{\dot{\mathcal{W}}_\sub{nf}(d,T_\sub{h},T_\sub{e})}{\dot{\mathcal{W}}_\sub{bb}(T_\sub{h},T_\sub{e})}, 
\end{equation}
which, taking into account the expansions to leading order in $\Delta T$ of both $\dot{\mathcal{W}}_\sub{nf}$ and $\dot{\mathcal{W}}_\sub{bb}$, can be written as
\begin{equation}
\Phi_0(d,T_\sub{e})=\frac{\hbar^2\omega_0^2}{4\sigma\kB T_\sub{e}^5}\frac{\ee^{\hbar\omega_0/(\kB T_\sub{e})}g_d(\omega_0)}{\left[\ee^{\hbar\omega_0/(\kB T_\sub{e})}-1\right]^2}. 
\end{equation}
Similarly, in order to compare efficiencies we introduce $\mathcal{R}(T_\sub{e})\equiv \lim_{T_\sub{h}\to T_\sub{e}} \bar{\eta}_\sub{nf}/\bar{\eta}_\sub{bb}$, which reads
\begin{equation}
\mathcal{R}(T_\sub{e})=\frac{\hbar\omega_0}{4\kB T_\sub{e}}\left[1-\ee^{-\hbar\omega_0/(\kB T_\sub{e})}\right]^{-1}.
\label{R}
\end{equation}
The condition $\mathcal{R}>1$ leads to a threshold frequency $\omega_\sub{th}=3.921 \kB T_\sub{e}/\hbar$~\cite{Latella:2014}. Thus, if $\omega_0<\omega_\sub{th}$, the conversion of near-field radiation cannot be more efficient than the conversion of blackbody radiation at small temperature difference.
We stress that $\omega_0$ is a property of the material. The functions $\Phi_0$ and $\mathcal{R}$ are related to each other, with the relation given by
\begin{equation}
\Phi_0(d,T_\sub{e})=\frac{\hbar \omega_0}{\sigma T_\sub{e}^4}g_d(\omega_0)n(\omega_0,T_\sub{e})\mathcal{R}(T_\sub{e}). 
\end{equation}

\begin{figure}
\includegraphics[scale=1]{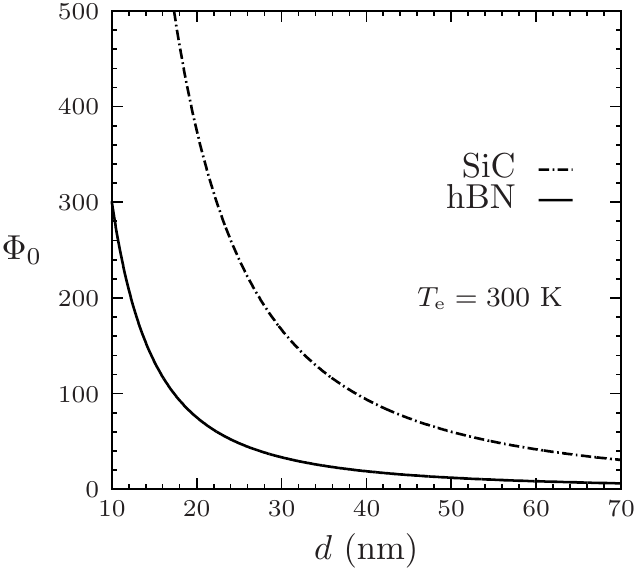}
\caption{Ratio $\Phi_0=\dot{\mathcal{W}}_\sub{nf}/\dot{\mathcal{W}}_\sub{bb}$ in the limit where the temperature of the hot source $T_\sub{e}$ approaches the environmental temperature $T_\sub{e}$ as a function of the gap width. In this asymptotic limit, the enhancement of the ideal work flux in the near-field regime as compared with that of blackbody radiation is shown for two different materials.}
\label{graph_Fi0}
\end{figure}

Of course, both the ideal work flux and the efficiency go to zero in the limit $T_\sub{h}\to T_\sub{e}$. However, the ratios $\Phi_0$ and $\mathcal{R}$ provide intrinsic information about the energy harvesting in the asymptotic limit of small temperature difference. In Fig.~\ref{graph_Fi0} we show $\Phi_0$ at $T_\sub{e}=300\;$K as a function of $d$ for SiC and hBN, where the optical data are taken from~\cite{Palik:1998} for the former material and from~\cite{Messina:2013} for the latter. In the figure it is clearly seen the enhancement in the work flux due to evanescent modes in the near-field regime; $\dot{\mathcal{W}}_\sub{nf}$ is considerably larger than $\dot{\mathcal{W}}_\sub{bb}$ at room temperature in the nanoscale.

\section{Conclusions}

We have studied the thermodynamics and energy harvesting of thermal radiation between two semi-infinite polar media separated by a nanoscale vacuum gap in the near-field regime.
Thermodynamic functions, such as energy and entropy fluxes, are constructed by computing the spectral flux of modes for the radiation between the surfaces of the materials. The thermodynamic approach and, in particular, the expression for the spectral flux of modes rest on the assumption that the macroscopic Maxwell equations hold at this scale. Thus, although being at the nanoscale, the system is considered to be macroscopic. According to this argument, in principle, there are no reasons to consider finite-size effects in the thermodynamic formalism, as occurs for small systems~\cite{Hill:2001} or for systems with long-range interactions~\cite{Latella:2013}. In small-scale systems, the transport could be affected by forces due to direct interactions between particles, hydrodynamic interactions or excluded volume effects, all these phenomena leading to an emergent dynamics that can be theoretically investigated~\cite{Reguera:2005}.  
In the present case, however, it is the considered thermodynamic system itself, i.e, the electromagnetic radiation, which shows a different behavior at the nanoscale in comparison to the behavior at macroscopic length scales. Such a difference appears here because evanescent modes play a very important role in the nanoscopic vacuum gap, while they are negligible for macroscopic separations.
Near-field thermodynamics is, therefore, described by the same thermodynamic relations of macroscopic systems, but with characteristic thermal coefficients containing the small-scale behavior. This formalism can generally be applied to systems that intrinsically modify their nature at non-macroscopic scales.
The properties of the system that depend on the working scale are codified in fundamental quantities such as the density of states or the spectral flux of modes.
These fundamental quantities are precisely those incorporated in the near-field thermodynamics, as we have shown here. 

Furthermore, using the near-monochromatic approximation~\cite{Rousseau:2012}, an analytical expression has been given for the maximum work flux that can be extracted from the radiation in the near field~\cite{Latella:2014}. This quantity has been compared with the one corresponding one to the blackbody regime and is shown to be considerably higher. 
An upper bound for the thermodynamic efficiency has also been studied.
Both maximum work flux and efficiency depend on the optical properties of the materials, and the explicit dependence on the frequency of the surface phonon polariton has been obtained.
Since the frequency of the surface waves depends on the choice of the material, our analysis highlights how the properties of the material influence the performance of the energy harvesting process.

\begin{acknowledgments}
I.L. acknowledges financial support through an FPI Scholarship (Grant No. BES-2012-054782) from the Spanish Government.
J.M.R. acknowledges financial support from Generalitat de Catalunya under program ICREA Academia.
This work was supported by the Spanish Government under Grant No. FIS2011-22603, and by CNPq and Funda\c{c}\~{a}o Arauc\'aria of the Brazilian Government. 
\end{acknowledgments}

\appendix*

\section{}
\label{appendix}
We now show that if the spectral flux of modes depends on $T$ but can be considered constant in a certain range of temperatures, equation~(\ref{delta_S}) for the variation of the entropy flux holds if $T_\sub{e}$ and $T_\sub{h}$ are in this range. Thus, consider now $\tilde{\varphi}=\tilde{\varphi}(\omega,T)$, so that the energy flux is given by
\begin{equation}
\dot{U}(T)=\int_0^\infty\dif \omega\; \hbar\omega n(\omega,T)\tilde{\varphi}(\omega,T).
\end{equation}
According to (\ref{entropy_flux}), we have
\begin{equation}
\dot{S}(T)=\int_0^\infty\dif\omega\ \hbar\omega\int_0^T \frac{\dif T'}{T'}\frac{\partial }{\partial T'}\left[n(\omega,T')\tilde{\varphi}(\omega,T')\right],
\end{equation}
and therefore $\Delta\dot{S}=\dot{S}(T_\sub{e})-\dot{S}(T_\sub{h})$ can be written as
\begin{equation}
\Delta\dot{S}=\int_0^\infty\dif\omega\ \hbar\omega\int_{T_\sub{h}}^{T_\sub{e}} \frac{\dif T'}{T'}\frac{\partial }{\partial T'}\left[n(\omega,T')\tilde{\varphi}(\omega,T')\right].
\label{delta_S_app}
\end{equation}
Our assumption here is that
\begin{equation}
\tilde{\varphi}(\omega,T)=
\begin{cases}
\varphi(\omega)&\mbox{if }T\in[T_a,T_b]\\
\phi(\omega,T)& \mbox{otherwise}
\end{cases},
\label{assumption}
\end{equation}
where $\phi(\omega,T)$ is an arbitrary function.
Hence, if both $T_\sub{e}$ and $T_\sub{h}$ lie on $[T_a,T_b]$, from (\ref{delta_S_app}) and (\ref{assumption}) we obtain
\begin{equation}
\Delta\dot{S}=\int_0^\infty\dif\omega\ \hbar\omega\varphi(\omega)\int_{T_\sub{h}}^{T_\sub{e}} \frac{\dif T'}{T'}\frac{\partial n(\omega,T')}{\partial T'},
\end{equation}
which does not depend on $\phi(\omega,T)$ and is equal to (\ref{delta_S}) by taking into account the definition of $m(\omega,T)$ given in~(\ref{m}).

\bibliography{mainbib}

\end{document}